\documentclass[a4paper,11pt]{article}
\pdfoutput=1 
\usepackage{jcappub}
\usepackage[T1]{fontenc}
\title{Angular distribution of low momentum atmospheric muons at ground level.}

\author[a,1]{I. Briki,\note{Corresponding author.}}
\author[a]{M. Mazouz}
\author[a]{L. Ghedira}

\affiliation[a]{Laboratoire de Physique Quantique et Statistique, Facult\'e des Sciences de Monastir, Universit\'e de Monastir, 5019 Monastir, Tunisia.}

\emailAdd{issabenmed@gmail.com}
\emailAdd{mazouz@jlab.org}
\emailAdd{lghedira7@gmail.com}

\abstract{
We report measurements of the angular distributions of low momentum atmospheric muons at 38 m above sea level for zenith angles $\theta$ between -60 and 60 degrees in the south-north direction. The muon detection was performed with two NaI(Tl) scintillation detectors mounted in coincidence. An adjustable lead thickness placed between the detectors allowed to select muons with a minimal momentum ranging from 0.3 to 0.9 GeV$/$c. The integrated and the differential muon flux were determined by analyzing the deposited energy spectra in the scintillators backed up by a Geant4 simulation of the experimental setup. The results are consistent with the $\cos^{n}(\theta)$ distribution in good agreement with the literature. These data contribute to fill the gap in the geomagnetic cutoff rigidity interval 8~GV$<P_c<$14~GV where no similar measurements were performed before. We found that $n=1.88-0.12~P_{\mu}^c$ in this domain of muon momenta cutoff  $P_{\mu}^c$<1 GeV$/$c. The present measurements are useful for many muon studies requiring an accurate integrated flux.}


\begin{document}
\maketitle
\flushbottom


\section{Introduction}
\label{sec.I}
 
Cosmic rays are a key part of the Universe. Their origin is connected to various astrophysical processes related to the evolution of stars.
Once accelerated out of their sources, cosmic rays spread across the interstellar medium before bombarding our planet. 
The charged component of these particles is mainly made up of high energy protons, helium nuclei and a tiny fraction of high Z nuclei.
Low energy particles are deflected by the Earth magnetic field while those with an energy exceeding few GeV enter our atmosphere and interact with its constituents to create hadronic and electromagnetic showers. Pions and kaons are the most abundant mesons in the hadronic shower and quickly decay into lighter leptons or $\gamma$ photons in the stratosphere. Atmospheric muons are the main charged particles which can reach the Earth surface with a mean energy of about 4 GeV. For such typical energies, muons interact with matter primarily through ionization, losing energy at a relatively constant rate of about 2 MeV~g$^{-1}$~cm$^{2}$~\cite{grieder2001cosmic}.  
Measuring the energy and angular distributions of secondary cosmic rays, and specially muons at various altitudes and latitudes, allows to shed light on many aspects of cosmic ray physics and remains an important topic nowadays. For example, atmospheric muons have been used as calibration sources for various experiments such as OPERA~\cite{mauri2020}, ANTARES~\cite{aguilar2010} and the Super-Kamiokande~\cite{abe2014}, investigating neutrinos flux and their oscillations.

In addition to being useful for understanding the interaction mechanisms of cosmic rays in the atmosphere, measuring muon flux provides a powerful tool for exploring various aspects of the world around us.
Indeed, these particles have found applications in different fields such as civil engineering~\cite{chaiwongkhot2018}, biology~\cite{Atri2001}, geophysics~\cite{lesparre2010geophysical, tanaka2016} and archaeology~\cite{morishima2017}. Most of these applications use the muography technique to construct a projection image of a target volume crossed by muons.
Muography was proved to be a useful tool even in exploring nuclear waste storage silos~\cite{jonkmans2013} and to construct a two-dimensional nuclear reactor map~\cite{perry2013} as well as for studying historic earthquakes recorded on ancient burial mounds~\cite{tanaka2020}. One of the muography main requirement is the accurate knowledge of the energy and angular distributions of atmospheric muon flux since it affects directly the imaging results. These distributions should also be determined in different geographic locations around the world and particularly at low energies ($<$1 GeV) where the muon intensities are sensitive to many factors (geomagnetic cutoff rigidity, altitude, average pressure and temperature, etc.)~\cite{Dorman2004}. These measurements become then of direct relevance in the validation of theoretical models or time-consuming simulation codes, which take into account such factors, if we want to use their flux predictions in the applications mentioned previously. It is worth mentioning that the determination of the energy and angular distributions of low energy atmospheric muons worldwide is also an important ingredient in a correct evaluation of the effective radiation dose taken up by the human body~\cite{sato2016} and in the irradiation studies of onboard electronics~\cite{kato2021,Sierawski2011}.


In this paper, we report the zenith angle distributions of low energy ($<$~1~GeV) atmospheric muons in Tunisia (region of Monastir: 35.76 N, 10.81 E) at sea level corresponding to a geomagnetic cutoff rigidity $P_c=$9.6 GV~\cite{copeland2020cari}.
To our knowledge, the only published muon flux data in the region of North Africa concern the vertical muon intensities that we measured in a recent work in the same location~\cite{issa}.
The results reported herein constitute then the first measurements of atmospheric muon angular distributions at sea level in this region of the world and for a geomagnetic cutoff rigidity close to 10 GV. The following section outlines the experimental setup and the performed configurations. Data analysis and the determination of muon flux based on a Monte-Carlo simulation fit to the experimental data are described in details in section~\ref{sec.III}. Finally, section~\ref{sec.IV} presents the obtained angular muon flux distributions and a comparison of our results with the literature.


\section{Experimental setup}
\label{sec.II}

\begin{figure}[h]
	\centering
	\includegraphics[width=1\textwidth]{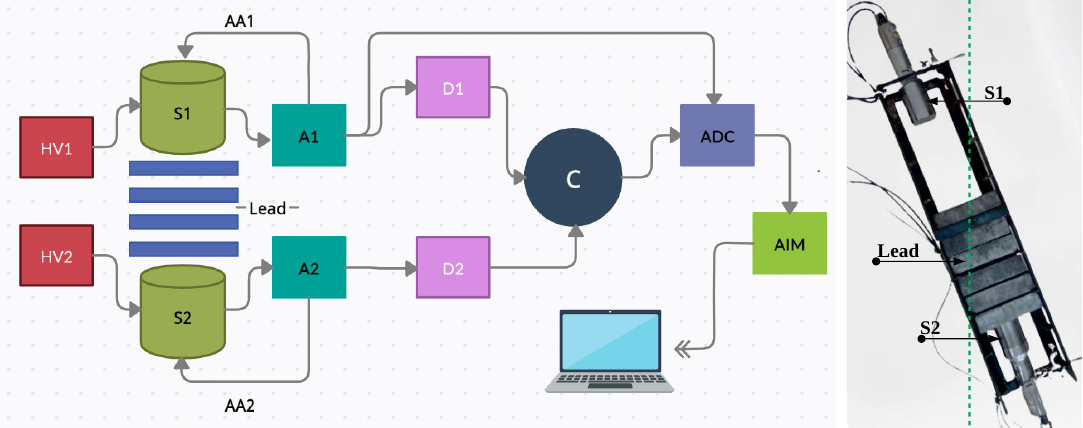}
	\caption{Schematic diagram of the experimental setup (left). A rotating holder, shown on the right, allows to align the two detectors (S1 and S2) and the lead layers in a given direction with respect to the vertical. HV1 and HV2 (AA1 and AA2) designate the high (low) voltage power supply for the PMTs (preamplifiers) of the S1 and S2 detectors. The passage of a particle through S1 or S2 produces an electrical signal which is amplified and shaped at a 0.5~$\mu s$ width by two main amplifiers (A1 and A2). Two constant fraction discriminators (D1 and D2) convert the signals passing a threshold, equivalent to 2 MeV energy deposit in S1 or S2, into logic gates. If a coincidence between two logic gates occurs within 1~$\mu s$ resolving time, the coincidence module (C) triggers the digital conversion of the detector signal amplitudes by the Analog to Digital Converter (ADC). The digital data are then stored in an Acquisition Interface Module (AIM) connected to a PC.}
	\label{fig1}
\end{figure}

The detection of atmospheric muons is performed by two coaxial cylindrical detectors separated by a distance equal to 64.5 cm.
The detectors are maintained by a rotating metallic holder in which a variable thickness of lead can be inserted. This ensures a stable alignment between the detectors and the lead layer when measurements at different angles are carried out. Each detector is a combination of a scintillator NaI(Tl) crystal (width~=~height~=~7.62~cm), a photomultiplier tube (PMT) and a preamplifier. The passage of a particle through one of the detectors produces an electrical signal whose amplitude is proportional to the energy deposited by the particle in the corresponding scintillator. As explained in figure~\ref{fig1}, the acquisition chain allows to record the detector signal amplitude when two energy losses higher than 2 MeV occur simultaneously in both detectors. This event can be identified later as an atmospheric muon being able to cross the two detectors and the lead thickness. The 2 MeV threshold serves to discard most of the surrounding background radiations~\cite{el2018determination, gilmore2008practical2}.


We performed three sets of measurements between March 2021 and January 2022. Each set corresponds to a specific thickness of lead (10, 30 and 50 cm), inserted between the two detectors, to select muons with three different minimal momenta $P_{\mu}^c$ according to muons-range tables in lead~\cite{groom2001muon}. For each lead thickness, eleven zenith angles $\theta$ have been covered between -60 and 60 degrees in the magnetic south-north direction respectively, where $\theta=0^{\circ}$ refers to the zenith direction. The acquisition time and the raw counts for each measurement are given in table~\ref{tab0}.
The experiment took place inside the physics department of the {\it Facult\'e des Sciences de Monastir} (38 m above sea level). The energy loss in the building (50 cm concrete) by muons before reaching the detectors was evaluated with a Monte-Carlo simulation and is about 197~MeV$/\cos(\theta)$. This energy loss is taken into account hereafter in order to present the results for open sky muons.

\begin{table}[h]
\centering
\begin{tabular}{|c|c|c|c|c|c|c|}
\hline
&\multicolumn{2}{c|}{10 cm Pb}& \multicolumn{2}{c|}{30 cm Pb} & \multicolumn{2}{c|}{ 50 cm Pb} \\
\hline
\hline
$\theta$& Time (h) & Counts &  Time (h) & Counts& Time (h)  &Counts\\ 
\hline
0& 134.83 & 3018 & 217.05 & 4219 &  194.03 & 3637 \\
\hline
20&  150.59 & 3120  & 183.72 & 3318  & 159.75 & 2938\\
\hline
30&  155.66 & 3074  & 186.41 & 3033  & 166.23 & 2879 \\ 
\hline 
40&  170.99 & 3045  &  167.78 & 2571 & 205.77 & 3000\\
\hline
50 & 220.56 & 3515  & 213.04 & 3123  & 261.64 & 3413\\ 
\hline
60 &  236.71 & 3230 &  267.17 & 3114 & 330.16 & 3725\\
\hline
-20 &  140.39 & 2934 & 159.41 & 3010  & 174.58 & 3228\\ 
\hline
-30&  165.90 & 3125 & 187.60 & 3182  & 162.43 & 2885 \\
\hline
-40 &  174.10 & 3003  &  203.05 & 3157 & 233.96 & 3137\\ 
\hline
-50 &  202.92 & 3135  & 210.05 & 3049  & 263.29 & 3246\\
\hline
-60 & 283.63 & 4044   & 240.09 & 3122  & 354.63 & 3514 \\
\hline     
     
\end{tabular}
\caption{Acquisition times and raw counts for the performed measurements.}

\label{tab0}
\end{table}


\section{Data analysis}
\label{sec.III}

\begin{figure}[h]
	\centering
	\includegraphics[width=0.8\textwidth]{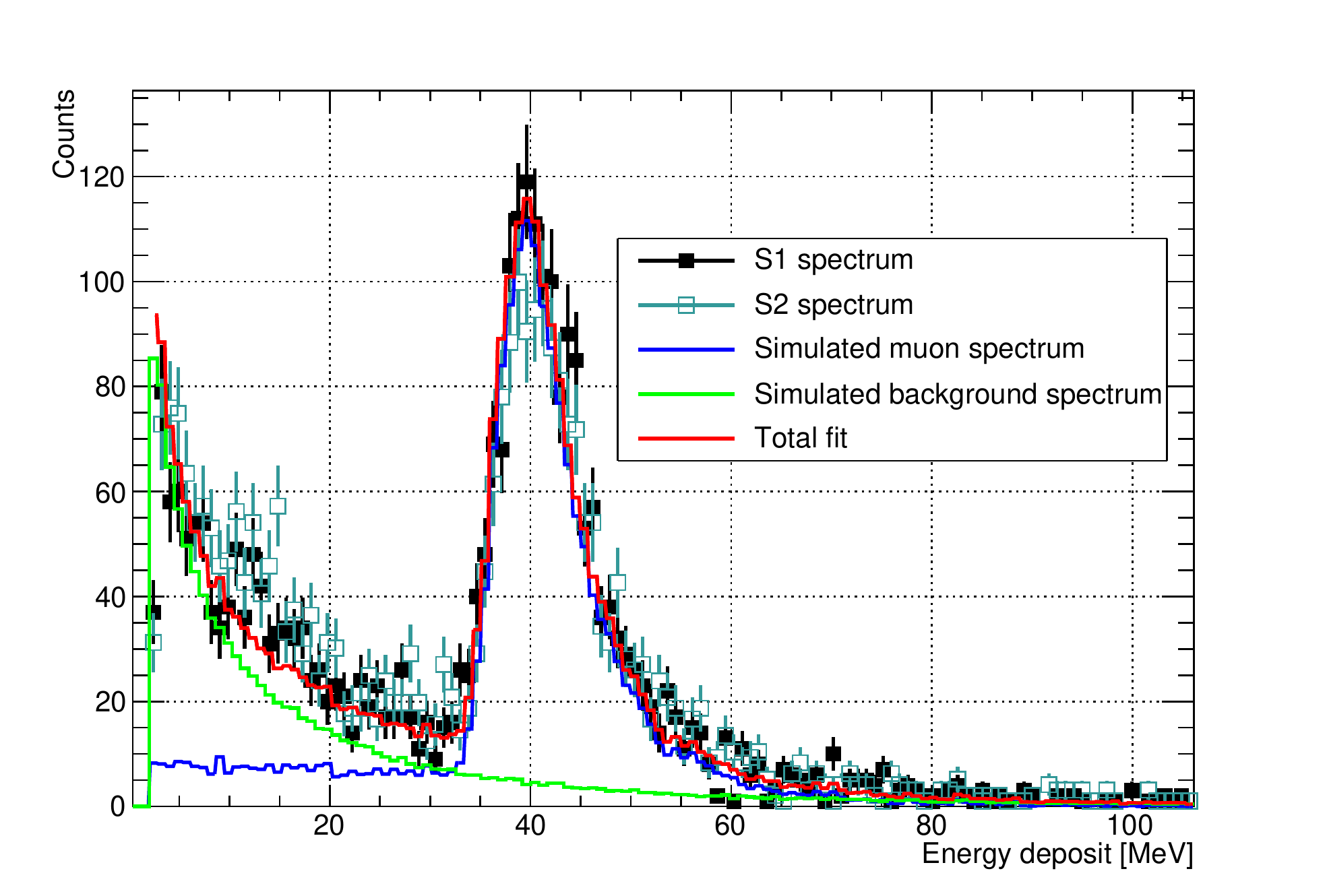}
	\caption{Experimental spectrum of the energy deposited in the S1 (black points) and S2  (cyan points) detectors corresponding to the configuration with 10 cm lead thickness and $\theta=-20$ degrees. Error bars represent the statistical uncertainties. The S1 experimental spectrum is fitted with the simulated background and muon contributions. The total fit is shown by the red curve and represents the sum of the optimal scaled background (in green) and muon (in blue) spectra.}
	\label{fig2}
\end{figure}

The energy deposited in a given detector is obtained by multiplying the recorded signal amplitude by a calibration coefficient. This coefficient is determined by performing dedicated runs with gamma-emitting radioactive sources of known energies. The energy calibration of both detectors and the counting rates were checked frequently to ensure a stable data acquisition during the experiment. Figure~\ref{fig2} shows an example of the spectra of the energy deposited in the S1 and S2 detectors for a given configuration. All the events in each spectrum are recorded when a coincidence between S1 and S2 occurs. The peak at the middle of each spectrum is due to muons passing through both detectors and the lead layer. Its mean position (about $\sim$40 MeV) is consistent with the energy loss by muons, at minimum ionization, along a distance equal to the vertical thickness of the scintillator. The lower edge of the spectra, called background hereafter, has an exponential shape and is compatible with the energy loss behavior of a photon or an electron coming from the cosmic ray shower or the interactions (decays) of muons~\cite{PREUSSE1994569, autran2018characterization}. The spectrum of the energy deposited in the S2 detector is similar to the S1 spectrum. The number of events in the peak is slightly lower than the one in the S1 spectrum because of the deflections of a few low energy muons in lead before crossing the edge of the S2 detector, resulting in smaller energy deposits in S2 for these muons.

\begin{figure}[h]
	\centering
	\includegraphics[width=1.04\textwidth]{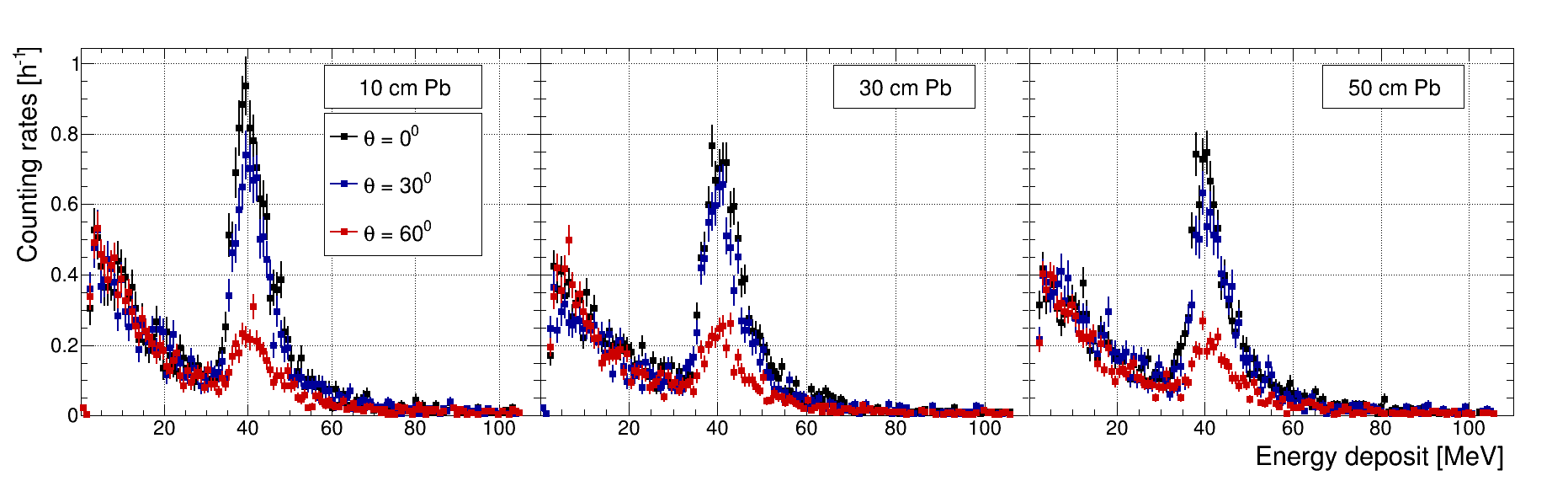}
	\caption{Experimental spectra of the energy deposited in the S1 detector for the three lead thickness configurations (10, 30 and 50 cm). The black, blue and red points correspond to the zenith angles 0, 30 and 60 degrees respectively. The number of events in each spectrum is normalized by the acquisition time. Error bars represent the statistical uncertainties.}
	\label{fig_bg}
\end{figure}

Figure~\ref{fig_bg} shows that the background contribution, for a given lead thickness, is independent of the zenith angle which suggests that the corresponding events are mainly due to fortuitous coincidences between two different particles detected in S1 and S2. However, The counting rates for the events in the peaks decrease as a function of the zenith angle which is consistent with muons passing through both detectors. At a given zenith angle, the signal-to-background ratio remains stable as a function of the lead thickness since both muons (signal) and background undergo a similar decrease with lead thickness. It is worth mentioning that the background contribution increases significantly when no lead plate is placed between the detectors. The signal-to-background ratio decreases in this case by at least a factor of 2 for all the zenith angles. The background contribution must be evaluated and subtracted from all the spectra in order to determine correctly the number of detected muons for each measurement. 

A Monte-Carlo simulation of the experimental device, based on the Geant4 package~\cite{agostinelli2003geant4}, was carried out to determine the expected shapes of muon and background spectra for each angle and lead thickness configuration. The CRY code~\cite{hagmann2007cosmic} was used in this simulation to generate cosmic rays at ground level. It has the advantage of producing realistic energy and angular distributions of cosmic ray particles (muons, photons, electrons, etc.) and allows to adjust important parameters such as the altitude, the latitude and the day of the experiment. The number of generated events was between 2.10$^8$ and 5.10$^8$ depending on the configuration. The same cuts, than those used experimentally on the detectors, have been then applied on the generated events in order to get simulated background and muon spectra. 

The determination of the number of detected muons $N_{\mu}$ for each configuration is obtained with the $\chi^2$ minimization technique : the experimental spectrum is fitted with the simulated background and muon spectra. The two free parameters of the fit are the scaling factor of the simulated muon spectrum and the scaling factor of the simulated background spectrum. The adjustment result is shown in figure~\ref{fig2} with the colored curves. The experimental number of muons is then deduced from the integral of the simulated muon spectrum shown in blue in figure~\ref{fig2}. Since the calibration of the detectors is done with gamma-emitting radioactive sources, which is only accurate between 0.5 and 3 MeV, a systematic uncertainty affects the position of the experimental muon peak relative to the simulated one. This uncertainty of the calibration at high deposited energies introduces a systematic uncertainty on $N_{\mu}$ which is evaluated by studying the variation of $N_{\mu}$ and the $\chi^2$ of the fit when changing the calibration coefficient inside a confidence interval. More details on the determination of $N_{\mu}$ and its systematic uncertainty can be found in our previous work~\cite{issa}.

The integral flux $\Phi_{I}$ of atmospheric muons for a given momentum cutoff $P_{\mu}^c$ is defined by the number of muons with a momentum $P_{\mu}>P_{\mu}^c$ arriving at the surface area $S$ of the face of the detector and following a direction $\theta$ inside a solid angle $\Delta\Omega$ during a time interval $\Delta t$~\cite{rossi1948interpretation}. Experimentally, it can be written as :

\begin{equation}\label{flux1}
\Phi_{I}(P_{\mu}^c,\theta) =  \frac{N_{\mu}}{\epsilon~S~\Delta\Omega~\Delta t}~~(m^{-2}~sr^{-1}~s^{-1})~,
\end{equation}

where $N_{\mu}(P_{\mu}^c,\theta)$ represents the number of detected muons and $\epsilon$ is a correction factor equal to the product of the intrinsic efficiency $\epsilon_{det}$ of the detectors ($>$99\%) and a geometrical correction $\epsilon_{geo}(P_{\mu}^c,\theta)$. The latter correction serves to increase the solid angle $\Delta\Omega=1.09~10^{-2}$~sr~\cite{gilly1978solid} between the two detectors in order to take into account multiple scattering events in the setup~\cite{Karmakar:1973qj} and the edge of the detector acceptance. This increase of the solid angle is between 6\%, in the vertical direction and for a lead thickness of 10 cm, and 14\% for the highest zenith angles and for a lead thickness of 50 cm. The Geant4 simulations coupled to the CRY generator, discussed previously, are used to determine the effective solid angle $\epsilon_{geo}\Delta\Omega$ for each configuration. Since the CRY integral flux $\Phi_{I}^{CRY}$ is known as well as the duration $\Delta t^{CRY}$ of the cosmic ray generation, the same experimental cuts applied to the simulated data allow to deduce the number of "detected" muons $N_{\mu}^{sim}$ in the simulation and thus $\epsilon_{geo}\Delta\Omega$ from eq.~\ref{flux1} :
\begin{equation}\label{fluxcry}
\epsilon_{geo}\Delta\Omega =  \frac{N_{\mu}^{sim}}{\Phi_{I}^{CRY}~S~\Delta t^{CRY}}~~.
\end{equation}


\section{Systematic uncertainties and stability check}
The statistical uncertainty on $\Phi_{I}$ originates from the statistical fluctuations on $N_{\mu}$ while its systematic uncertainty has many origins. As mentioned previously, the first ones come from the calibration-induced systematic uncertainty on $N_{\mu}$ and the 1\% uncertainty on $\epsilon_{det}$. The relative systematic uncertainties, induced by the calibration, vary between 1.5\% and 3\% for all the measurements. The determination of the effective solid angle $\epsilon_{geo}\Delta\Omega$ with the simulation (eq.~\ref{fluxcry}) introduces an additional systematic uncertainty which is related to the statistical fluctuations of $N_{\mu}^{sim}$. The latter depends on the simulated configuration and is about 1.5\% at most for the highest zenith angles where, despite the generation of 5.10$^8$ primary events in the simulation, we end up with a few thousand events for $N_{\mu}^{sim}$. Another systematic uncertainty could arise from an experimental misalignment of the detector axis and the precise knowledge of the distance between the active areas of the scintillators. This uncertainty which affects directly the real solid angle, and thus $\Phi_{I}$, should not exceed 1\% according to the precision of our pointing instruments. All the mentioned systematic uncertainties are added quadratically to obtain the total systematic uncertainty for each measurement.

\begin{figure}[h]
	\centering
	\includegraphics[width=0.49\textwidth]{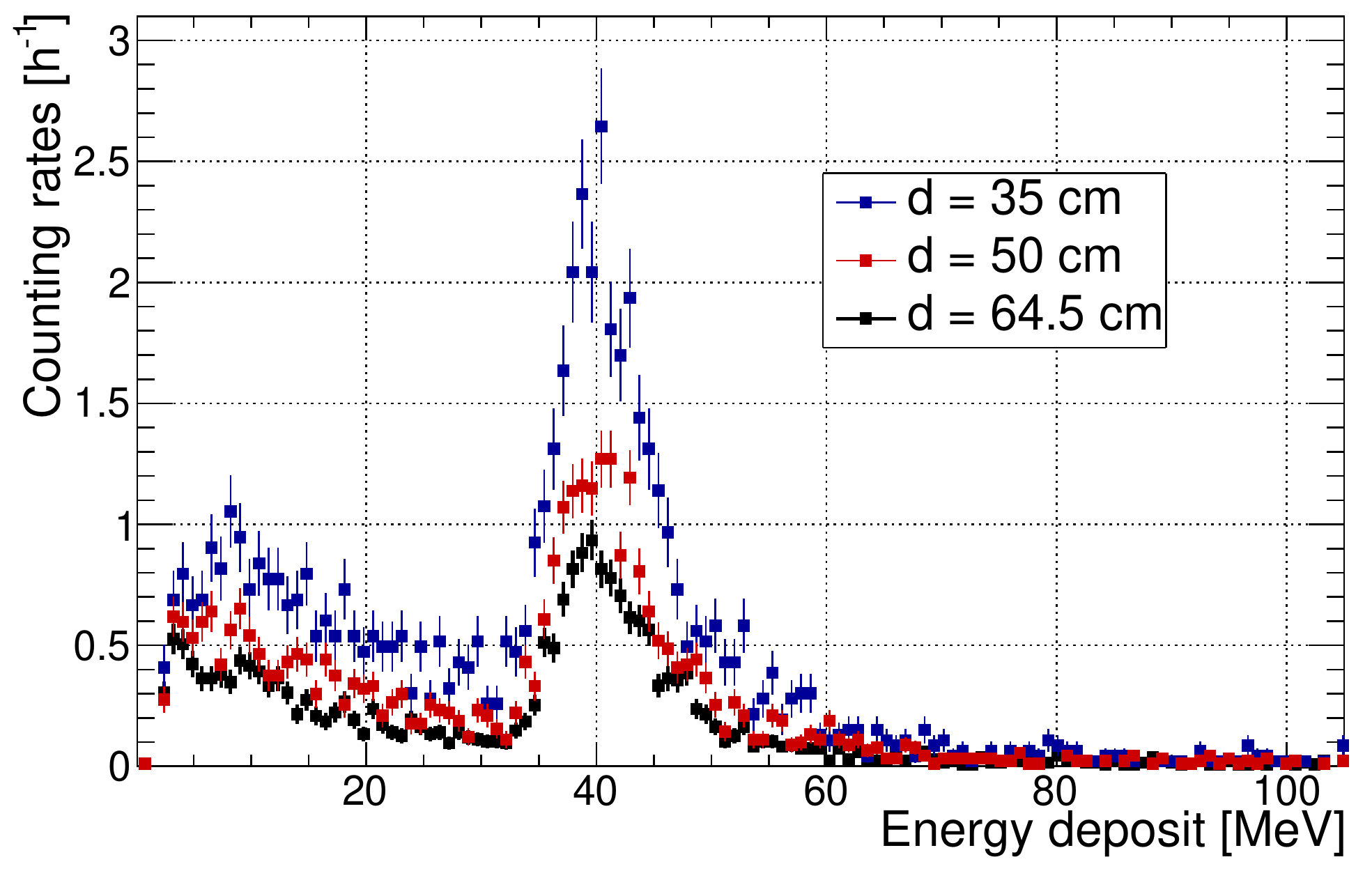}
	\includegraphics[width=0.49\textwidth]{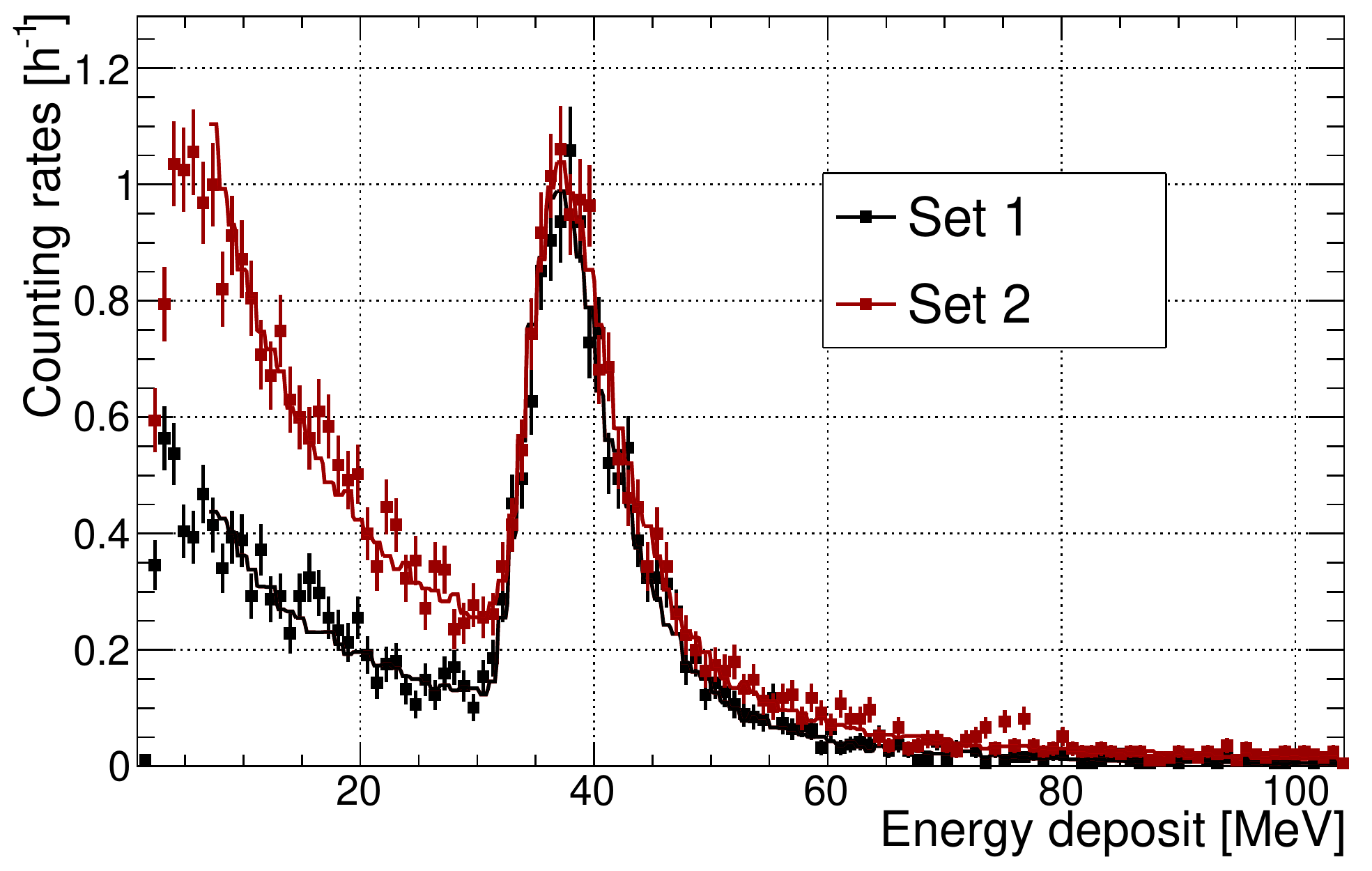}
	\caption{Spectra of the energy deposited in S1 in the vertical direction for the configuration with 10 cm lead thickness. Each spectrum in the left represents a specific distance d between the S1 and S2 detectors as indicated in the legend. Each spectrum on the right corresponds to a different position of the 10 cm lead thickness when d$=$64.5 cm. The lead block in Set 1 is placed just above the S2 detector, while in Set 2 it was raised 35 cm above S2. The two solid lines in the right figure represent the total fit of the two experimental spectra with the simulated ones.}
	\label{distance}
\end{figure}

As stated in the previous section, the determination of muon numbers is obtained by adjusting the experimental spectra by the simulated muon and background spectra. Even if the interaction cross sections of the background particles with the experimental setup are well known at the studied energy range, the shape of the simulated background spectra partly rely on the energy and angular distributions of each kind of background particles generated by CRY. This model-dependent background shape can then introduce a possible systematic uncertainty when extracting $N_{\mu}$. The distributions of muons used in the simulation are also model-dependent which is also likely to introduce uncertainties in the acceptance calculations (eq.~\ref{fluxcry}). To estimate the model dependency effect, we have performed three measurements for a given configuration with three different inter detector distances. The S1 energy deposit spectrum for each measurement is represented in figure~\ref{distance} (left). As expected, both muons and background contributions vary as a function of the distance between the detectors. This allows a modification of the solid angle and offers the possibility to check the modeling of the two contributions and our calculation of the solid angle by comparing the corresponding muon integral flux. After simulating the three configurations and applying the same analysis, exposed in section~\ref{sec.III}, we found the values reported in table~\ref{table01}. The three flux are compatible with each other within statistical uncertainties and the difference between the extreme values remains lower than the quoted systematic uncertainties. They are also compatible with the integral flux determined in our previous work~\cite{issa} ($\Phi_{I}=67.55\pm 2.06_{-2.71}^{+1.36}$) at the same lead thickness, but where the inter detector distance was 81 cm. This comparison, as well as the relatively good $\chi^2/ndf$ (about 1.4 in average) of the fits for all the experimental spectra, suggests that Geant4-CRY modeling of the background and muon spectra is quite realistic. The reasonable agreement between the CRY angular distributions of muon flux and our results, presented in the next section, also supports this assertion. The systematic uncertainties introduced by the modeling of the background and the muons should then not exceed those cited.

\begin{table}[h]
\centering
\begin{tabular}{|c|l|}
\hline
Distance (cm)  &$\Phi_{I} \pm \text{stat.}\pm \text{sys.}$   \\
\hline
  $64.5$    &  $ 70.58  \pm 1.93_{-2.91}^{+1.41} $ \\

  $50.0$    &  $ 67.87  \pm 1.84_{-3.00}^{+1.36} $ \\

  $35.0$    &  $ 69.92  \pm 1.87_{-2.58}^{+1.40} $  \\
\hline
\end{tabular}
\caption{Integral flux ($m^{-2}~sr^{-1}~s^{-1}$) for three different inter detector distances.}
\label{table01}
\end{table}

As mentioned earlier, the solid angle increase by the factor $\epsilon_{geo}$ is mainly due to scattering effects in the lead layer.  We have performed two additional sets of measurements for the 10 cm lead thickness configuration, but with two different positions of the lead block, to check if the multiple scattering effects are well taken into account in the Geant4 simulation and if the quoted systematic uncertainty (1.5\%) on $\epsilon_{geo}\Delta\Omega$ is realistic. In Set 1, the lead block is positioned just above the S2 detector, while in Set 2 it has been raised by a distance of 35 cm relative to S2. As shown in Figure~\ref{distance} (right), the energy deposit spectra for these measurements differ significantly only in the background region. Indeed, the S2 detector becomes less shielded by the lead block in Set 2, which increases the number of particles reaching S2 and leads to additional fortuitous coincidences. However, the counting rate of the muons crossing both detectors is almost the same for the two measurements and the corresponding integral fluxes differ by less than 1.5\% ($\Phi_{I} = 74.28 \pm  1.75_{-2.79}^{+1.49}~m^{-2}sr^{-1}s^{-1}$ for Set 1 and $\Phi_{I} = 75.39 \pm  1.93_{-3.03}^{+1.54}~m^{-2}sr^{-1}s^{-1}$ for Set 2). This comparison, as well as the good reproduction of the experimental spectra by the simulation, indicate that the uncertainty on the scattering effect modeling should not exceed that quoted.


\section{Results and discussion}
\label{sec.IV}

\begin{figure}[h]
	\centering
	\includegraphics[width=.9\textwidth]{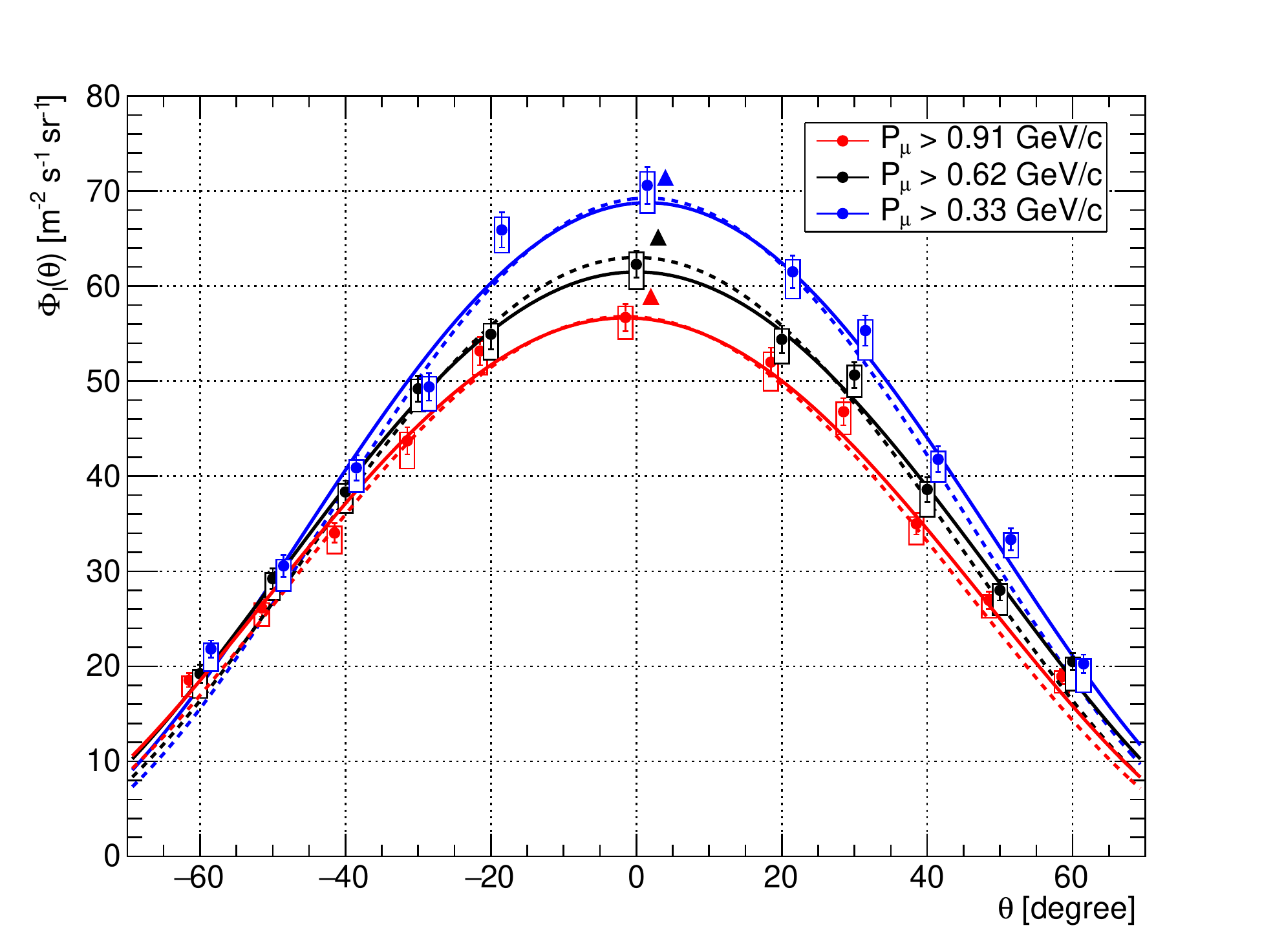}
	\caption{The zenith angle distributions of open sky muons for three lead thicknesses (10 cm in blue, 30 cm in black and 50 cm in red). The corresponding muon minimal momenta are indicated in the legend. The Error bars represent the statistical uncertainties while the white boxes around the points show the total systematic uncertainties. The solid and dashed lines represent the data fit with Eq.~\ref{flux3} and the CRY predictions respectively. The blue and red points and lines, are shifted by $ \pm 1.5^{\circ}$ relative to the black ones for clarity.
    The three triangles represent the vertical intensities measured in our previous work~\cite{issa}. They are compatible with the present measurements of $\Phi_{I}(0^{\circ})$, within error bars, and are shifted around $\theta=3^{\circ}$ for clarity.}
	\label{fig3}
\end{figure}

Figure~\ref{fig3} shows the integral flux angular distributions of open sky muons for the three lead thickness configurations. These distributions reflect a convolution of the production spectrum of muons, their path length in the atmosphere and their decay probabilities. Several formulas have been proposed in the literature to calculate the column density for inclined muon trajectories~\cite{grieder2001cosmic,shukla2018}, relating the integral flux at a given zenith angle $\Phi_{I}(\theta)$ to the vertical intensity $\Phi_{I}(0^{\circ})$ in curved Earth's atmosphere. For zenith angles not too large, the approximation of a flat Earth usually describes well the experimental data and leads to the simple expression~\cite{greisen1942intensities}:

\begin{equation}\label{flux3}
\Phi_{I}(\theta) = \Phi_{I}(0^{\circ})~ \cos^{n}(\theta)~~,
\end{equation}

where $n$ is a function of the lower muon momentum cutoff $P_{\mu}^c$. The data represented in figure~\ref{fig3} are fitted by eq.~\eqref{flux3}, using  $\Phi_{I}(0^{\circ})$ and $n$ as free parameters.
The MINUIT package~\cite{minuit} of the ROOT program~\cite{rootfit} was used to perform the fit and the estimation of the parameter errors. The $\chi^2/ndf$ of the three fits, where $ndf=9$ is the number of degrees of freedom, are below 1 when the quadratic sum of the statistical and systematic uncertainties is attributed to the integral flux errors of figure~\ref{fig3}.
The fitted values of $\Phi_{I}(0^{\circ})$ and $n$ are summarized in table~\ref{tab1} in comparison with the results of other experiments and CRY predictions. The geomagnetic latitude and the altitude are the main factors affecting the vertical intensities. Indeed, $\Phi_{I}(0^{\circ})$ decreases as moving closer to the equator due to an increasing effective vertical cutoff rigidity $P_c$. The higher $\Phi_{I}(0^{\circ})$ value of~\cite{riggi2020} is only due to the high altitude of the measurement. As expected, the present measurements of $\Phi_{I}(0^{\circ})$ are compatible with the values reported in our previous work~\cite{issa}, as can be seen in figure~\ref{fig3}. They are also coherent with the values reported in the literature when their variation as a function of $P_c$, $P_{\mu}^c$ and the altitude is considered.

\begin{table}[h]
	\centering
	\begin{tabular}{|l|c|c|c|c|c|c|l|}
		\hline
		Authors &  $ P_{c} $&Alt.& $P_{\mu}$&$n$ value &$\Phi_{I}(0^{\circ})$ \\
		& $(GV)$ &(m)&  $(GeV/c)$&&$(m^{2}~sr~s)^{-1}$\\
		\hline
		Pethuraj et al.
		\cite{pethuraj2017measurement}&17.6&160&$\geq$0.11&2.00$\pm$0.16&70.07$\pm$5.26\\
		\hline
		Sogarwal et al.\cite{sogarwal2022}&16.38&SL&$\geq$0.25&2.10$\pm$0.25&66.70$\pm$1.54\\
		\hline
		S. Pal et al.\cite{pal2012measurement}&16&SL&$\geq$0.28&2.15$\pm$0.01&62.17$\pm$0.05\\
		\hline
		Bhattacharyya \cite{bhattacharyya1974}&14&24&$\geq$0.4&1.91$\pm$0.1&-\\
		&&&$\geq$1.&1.85$\pm$0.11&-\\
		\hline
		Arneodo et al.\cite{arneodo2019measurement}&14&SL&$\geq$0.04&1.91$\pm$0.18&75.4$\pm$1.4\\
		\hline
		Present data&9.6&38&$\geq$0.33&1.82$\pm$0.11&68.77$\pm$1.94\\
		&&&$\geq$0.62&1.72$\pm$0.10&61.49$\pm$1.44\\
		&&&$\geq$0.91&1.72$\pm$0.10&56.66$\pm$1.60\\
		\hline
		CRY \cite{hagmann2007cosmic}&9.6&SL&$\geq$0.33&2.02&69.26\\
		&&&$\geq$0.62&1.95&63.02\\
		&&&$\geq$0.91&1.87&56.80\\
		\hline
		Riggi et al. \cite{riggi2020}&8&3100&$\geq$0.2&1.83$\pm$0.13&83.$\pm$8\\
		\hline
		Judge and Nash \cite{judge1965}&2.5&SL&$\geq$0.7&1.96$\pm$0.22&-\\
		\hline
	\end{tabular}
	\caption{A compilation of $\Phi_{I}(0^{\circ})$ and $n$ from different measurements at low cutoff muon momenta. }
	\label{tab1}
\end{table}

\begin{figure}[h]
	\centering
	\includegraphics[width=.85\textwidth]{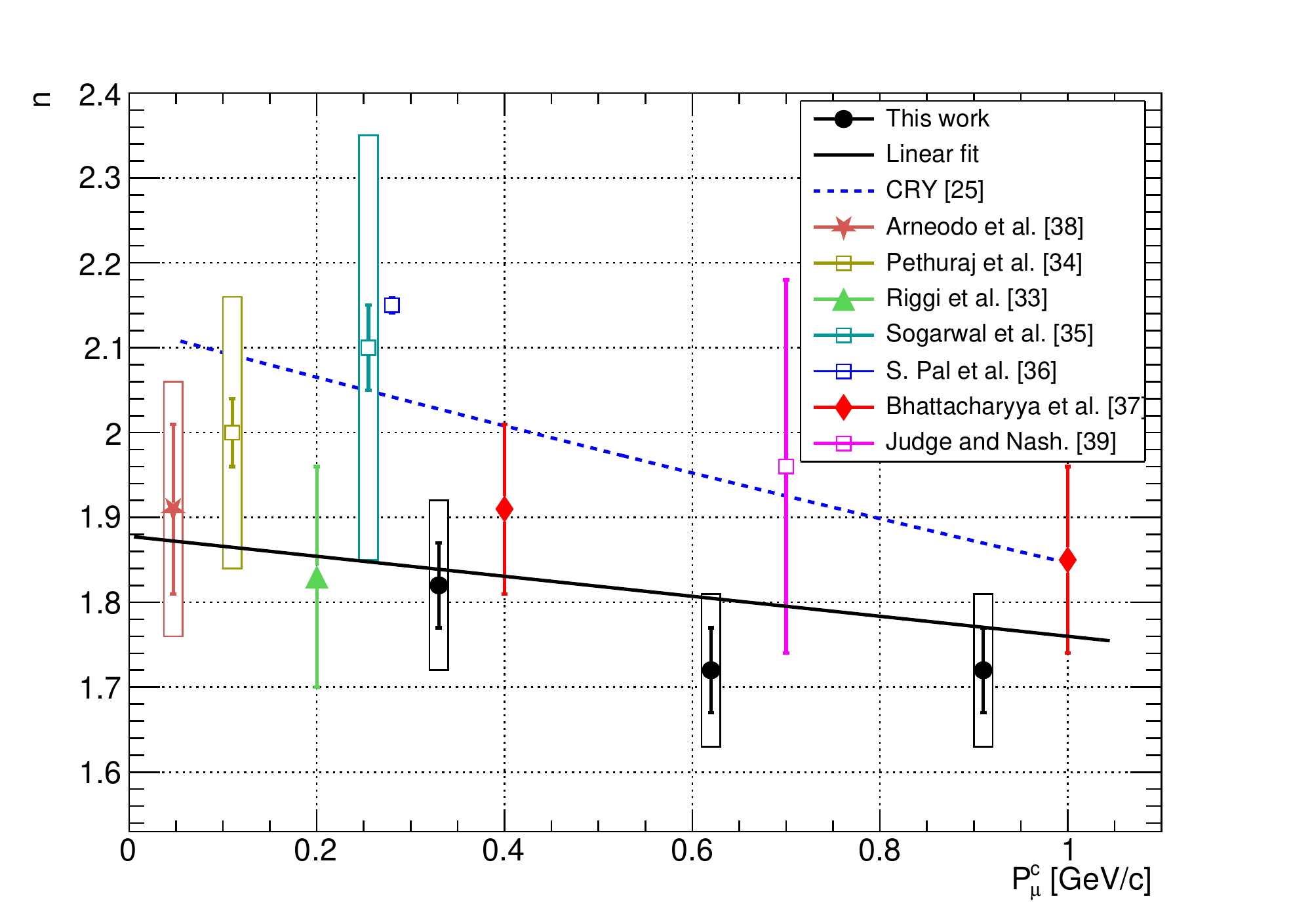}
	\caption{The $n$ exponent as a function of the muon momentum cutoff $P_{\mu}^c$. The black points show our extracted values from a fit of the angular distributions of figure~\ref{fig3}. The error bars (white boxes) represent the statistical (total systematic) uncertainties. Colored markers are similar results from the literature as indicated in the legend. The dashed line represents the CRY predictions while the solid line is a linear fit of the seven data points at the closest geomagnetic latitudes to ours (8~GV$\leq P_c \leq$~14~GV). Data with square markers are not included in the fit.}
	\label{fig4}
\end{figure}

Table~\ref{tab1} shows that the $n$ values are overall around 2 which gives the popular $\cos^{2}(\theta)$ law widely-used to get fast estimations of the angular distribution of muon flux.  At altitudes close to the ground level and at a given cutoff rigidity, the exponent $n$ is expected to decrease with increasing $P_{\mu}^c$ since the angular distribution tends to be flat then approaches a $\sec(\theta)$ distribution at increasing muon momenta~\cite{pdg2020}. Unfortunately, the dependence of $n$ on the cutoff rigidity $P_c$ is not yet very clear due to insufficient experimental results and large uncertainties on the reported values. Indeed, table~\ref{tab1} suggests a discrepancy between the reported $n$ values at $P_c\geq16$ GV which are all above 2 and the ones at $P_c<16$ GV which are all below 2. For example, this discrepancy is more than two standard deviations between the value of~\cite{pal2012measurement} and the values of~\cite{bhattacharyya1974} and~\cite{riggi2020}. As a precaution and to avoid an eventual $P_c$ dependence, we have chosen to fit linearly our extracted $n$ values along with the ones reported in table~\ref{tab1}, at the closest geomagnetic latitudes to ours (8~GV$\leq P_c \leq$~14~GV), as shown in figure~\ref{fig4}. This yields $n=1.88-0.12~P_{\mu}^c$ for open sky muons with a minimal momentum $P_{\mu}^c$ below 1 GeV$/$c. If the $P_{\mu}^c$ dependence of $n$ is neglected within the studied range, a fit of the same data points by a constant gives $n=1.81\pm0.04$. However, the linear expression of $n$ reproduces better the experimental data and should be used for the applications requiring an accurate knowledge of the muon integral flux.

One can see from figure~\ref{fig4} that the CRY Monte-Carlo simulation model overestimates the $n$ exponent by more than 10\% even if it reproduces globally the energy~\cite{issa} and angular distributions in figure~\ref{fig3}. This can be explained by the flat atmosphere model used in CRY which overestimates the muon path length for large zenith angles and thus underestimates the corresponding muon flux. The CRY integral flux decrease then slightly more rapidly at large zenith angles than the measured ones, as can be seen in figure~\ref{fig3}, resulting in higher $n$ values.\\

\begin{figure}[h]
	\centering
	\includegraphics[width=.8\textwidth]{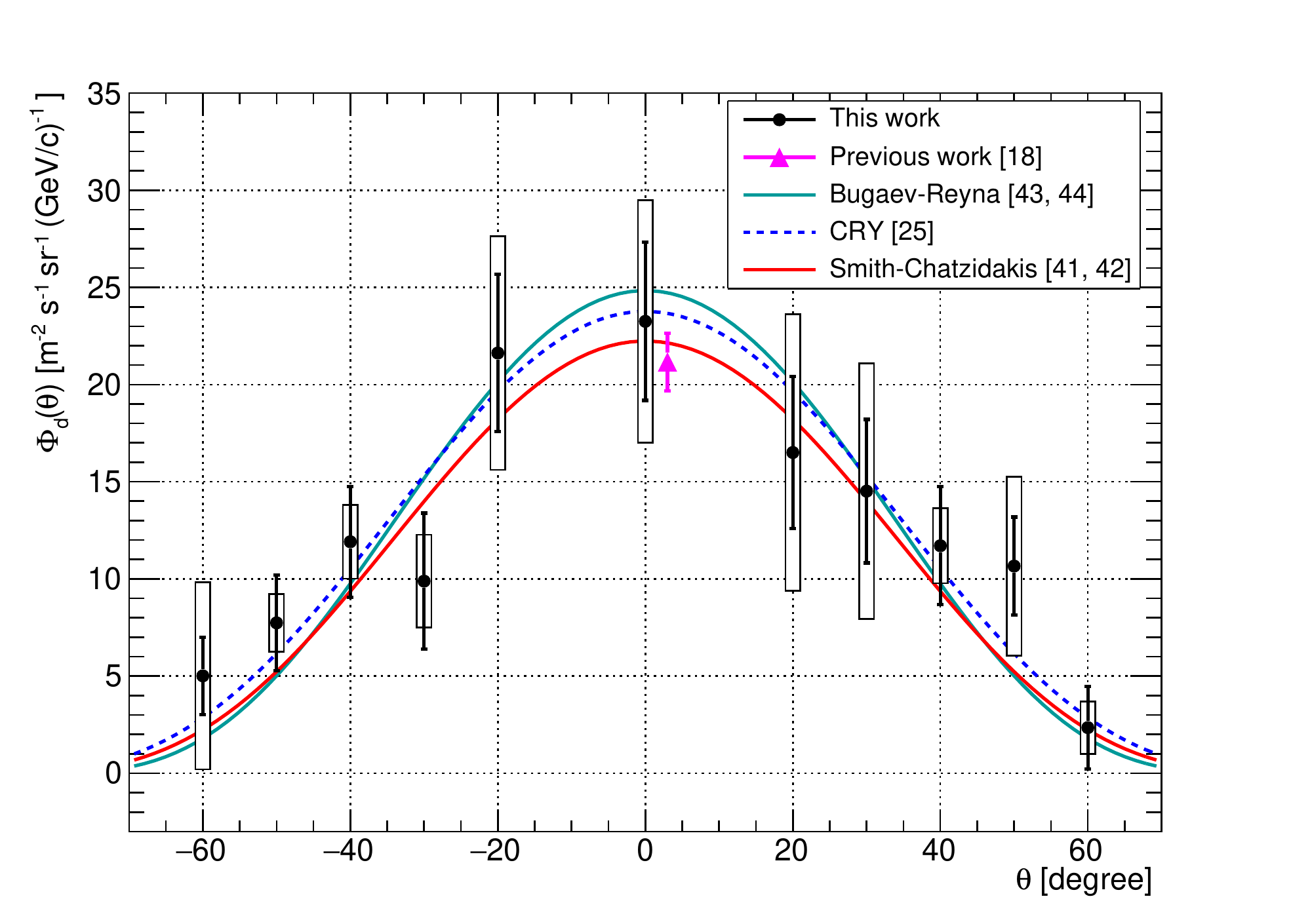}
	\caption{The angular distribution of the differential muon flux at ground level for $P_{\mu}\approx$0.61 GeV/c. The error bars represent the statistical uncertainties while the white boxes around the points show the total systematic uncertainties. The dashed line represents the CRY model predictions and the solid lines show the results of two parametric analytical models as indicated in the legend. The magenta triangle is the vertical differential flux from our previous work~\cite{issa}, it is shifted to the right by $3^{\circ}$ for clarity.}
	\label{fig5}
\end{figure}

The differential muon flux at a given muon momentum $P_{\mu}$ is defined by:
\begin{equation}
\Phi_{d}(\theta,P_{\mu})=\frac{d\Phi_{I}(\theta,P_{\mu})}{dP_{\mu}}~~(m^{-2}~sr^{-1}~s^{-1}~(GeV/c)^{-1})~,
\end{equation}
and can be determined from the $P_{\mu}^c$ dependence of the integral flux. As mentioned in section~\ref{sec.II}, three measurements of $\Phi_{I}(\theta,P_{\mu}^c)$ were performed for each zenith angle, corresponding to three $P_{\mu}^c$ values. The differential flux is deduced from the slope of a linear fit adjusting $\Phi_{I}(\theta,P_{\mu}^c)$ as a function of $P_{\mu}^c$. The obtained results are shown in figure~\ref{fig5}. The limited number of lead thickness configurations results on larger uncertainties in the present results relative to our previous measurement of the vertical differential flux where a larger range of $P_{\mu}^c$ was covered. Figure~\ref{fig5} shows that our two measurements of the vertical differential flux are consistent with each other. The CRY model reproduces well the shape of the angular distribution, within experimental uncertainties, for an average muon momentum $P_{\mu}\approx$0.61~GeV/c. At such low momenta, the muon path length in the atmosphere is relatively short and the flat Earth approximation used by CRY does not seem to have any effect on the differential muon flux for zenith angles lower than 60$^{\circ}$. The results are also in good agreement with two parametric analytical models, Smith-Chatzidakis~\cite{Smith1959,CHATZIDAKIS201533} and Bugaev-Reyna~\cite{Bugaev1998,Reyna2006}, which are among the best models reproducing the experimental data at $\theta=0^{\circ}$ and $\theta=75^{\circ}$ according to~\cite{NSu2021}.

\section{Conclusion} 
\label{sec.VII}
The angular distribution of atmospheric muons at ground level was measured for the first time in Tunisia at a geomagnetic cutoff rigidity not studied before. A variable lead thickness was placed between the two detectors to select only muons above a given momentum threshold. Our results show globally a good agreement with the results reported in the literature and the expected flux from a Monte-Carlo simulation code where the latitude, the altitude and the muon cutoff momentum are taken into account. These measurements could shed light on the variation of the $n$ exponent in the $\cos^{n}(\theta)$ distribution law when combined with other measurements at different locations and muon energies. The reported results represent an important input for many possible muon-based studies in our region, notably for the muography of historical and archaeological sites and a more accurate evaluation of the cosmic-ray irradiation doses.



\bibliographystyle{JHEP}
\bibliography{ref}

\end{document}